\documentclass[conference]{IEEEtran}
\usepackage{amssymb,amsfonts,amsmath, bbm,amstext}
\usepackage{graphicx,times,latexsym,makeidx,xspace,url,comment,subfigure,cite, multirow}
\usepackage{theorem}
\usepackage{color}

\newtheorem{theorem}{Theorem}[section]
\newtheorem{prop}[theorem]{Proposition}

\newtheorem{obs}[theorem]{Observation}

\newcommand{\I}[1]{{\mathbbm{1}_{#1}}}

\newcommand{\secR}[1]{Sec.~\ref{sec:#1}}

\newcommand{\figR}[1]{Fig.~\ref{fig:#1}}
\newcommand{\figLC}[2]{
        \caption{#2}
        \label{fig:#1}
        \vspace{-5pt}
}

\newcommand{\reno}[0]{TCP Reno}

\newcommand{\btledbat}[0]{LEDBAT}
\newcommand{\fledbat}[0]{fLEDBAT}

\begin{document}

\title{Rethinking low extra delay background\\ transport protocols}
\author{\IEEEauthorblockN{
Giovanna Carofiglio\IEEEauthorrefmark{1},
Luca Muscariello\IEEEauthorrefmark{2},
Dario Rossi\IEEEauthorrefmark{3},
Claudio Testa\IEEEauthorrefmark{3} and
Silvio Valenti\IEEEauthorrefmark{3}
\\}
\IEEEauthorblockA{
\IEEEauthorrefmark{1} Bell Labs, Alcatel-Lucent, France, giovanna.carofiglio@alcatel-lucent.com\\}
\IEEEauthorblockA{
\IEEEauthorrefmark{2} Orange Labs, France Telecom, France, luca.muscariello@orange-ftgroup.com\\}
\IEEEauthorblockA{
\IEEEauthorrefmark{3} Telecom ParisTech, France, firstname.lastname@enst.fr}
}

\maketitle

\begin{abstract}
BitTorrent has recently introduced LEDBAT, a novel application-layer congestion control protocol for data exchange. The protocol design starts from the assumption that network bottlenecks are at the access of the network, and that thus user traffic competes creating self-inducing congestion. To relieve from this phenomenon, LEDBAT is designed to quickly infer that self-induced congestion is approaching (by detecting  relative changes of the one-way delay in the transmission path), and to react by reducing the sending rate prior that congestion occurs. Prior work has however shown LEDBAT to be affected by a latecomer advantage, where newly arriving connections can starve already existing flows.
In this work, we propose modifications to the congestion window update mechanism of the LEDBAT protocol that aim at solving this issue, guaranteeing thus intra-protocol fairness and efficiency.
Closed-form expressions for the stationary throughput and queue occupancy are provided via a fluid model, whose accuracy is confirmed by means of ns2 packet level simulations.
Our results show that the proposed change can effective solve the latecomer issue, without affecting the other original LEDBAT goals at the same time.
\end{abstract}

\section{Introduction}\label{sec:intro}

BitTorrent, which certainly needs no introduction due to its tremendous popularity, has recently developed a novel application-layer congestion control  protocol for data exchange. While congestion control is a long studied subject, the refreshing ingredient is in this case the reasonable assumption that the bottleneck is likely placed at the access of the network (e.g., at the ADSL modem line), which means that congestion is therefore typically self-induced by concurrent traffic generated by the user (e.g., BitTorrent transfers in parallel with Skype call and Web browsing).

This novel protocol, named \btledbat\ after Low Extra Delay Background Transfer, is designed to solve this issue and targets (i) efficient but (ii) low priority transfers. When \btledbat\ flows have the exclusive use of the bottleneck resources, they fully exploit the available capacity. When instead other transfers --such as VoIP, gaming, Web or other TCP flows-- are ongoing, \btledbat\ flows back off to avoid harming the performance of interactive traffic.

To attain the efficiency aim, \btledbat\ flows need to create queuing, as otherwise the capacity would not be fully utilized. At the same time, due to the low-priority aim, the amount of extra queuing delay caused by \btledbat\ flows should be small enough to avoid hurting the interactive traffic -- hence the protocol name.

 \btledbat\ has been defined as an IETF draft~\cite{ledbat_draft} (which focuses more on the algorithmic aspects) and as a BitTorrent Enhancement Proposal~\cite{bep29} (that instead focuses more on the UDP framing).
 Since version 2.0.1 released on April 2010, the protocol has become BitTorrent default congestion control protocol, replacing thus TCP\footnote{Notice that the protocol has been christened as \btledbat\ in the IETF community, and as uTP in the BEP community: in this paper, to avoid ambiguity, we use its IETF name.}.

While previous research\cite{pam10,icccn10,globecom10,lcn10,gordon2010iccnt,cohen10iptps} on \btledbat\ has shown its potential, it also has highlighted some limits -- above all, a fairness issue that arises in case of backlogged flows, where latecomer flows take over the bottleneck resource, starving the first-comers. Though not critical as far as the low-priority goal is concerned, unfairness may affect other aspect of the protocol (e.g., interaction with peer selection and tit-for-that) and deserves therefore further attention.

The main contribution of this work is to propose a modification to the \btledbat\ congestion control that, leaving untouched the other goals, resolves the fairness issue.
Throughout this paper, we make use of several complementary techniques to study our proposal. First, we use an active testbed methodology to show the fairness issue in current BitTorrent. Then, we develop a fluid model describing the system dynamics in the case that several \btledbat\ flows shares the same bottleneck. Analytical solution of the model gives us useful insight on the regime performance of the modified \btledbat\ protocol. The model is complemented with numerical solutions that allow to
grasp the transient phase as well. Finally, we use \texttt{ns2} packet-level simulation to evaluate \btledbat\ performance under more realistic scenarios.

The reminder of this paper is organized as follows. Related work and motivations are covered in \secR{related} and \secR{motivation} respectively. Our proposed modification to \btledbat\ is presented in \secR{model}, along with the fluid model and its analytical solution. Moreover, the same section also reports a comparison between the numerical solution of the fluid model and the packet level simulation in simple scenarios. Finally, more realistic network conditions are tackled by means of simulation in \secR{simulation}, assessing the importance of the traffic model (e.g., backlogged vs chunk-based transfers), of the protocol tuning (i.e., sensitivity analysis to protocol parameters) and of the scenario realism (e.g., homogeneous vs heterogeneous environment).  Conclusive remarks are reported in \secR{outro}.

\section{Related Work}\label{sec:related}

Congestion control studies on the Internet date back to \cite{jacobson88tcp} and it would be therefore very hard, other than out-of-scope, to provide a full review of the existing literature here. Still, a couple of references are worth citing as they share \btledbat\ low-priority spirit~\cite{tcp_nice,tcp_lp,tcp_4cp,tcp_key}.
Similarly, BitTorrent has not only become a largely popular application among its users, but it has become a popular research subject as well. At the same time, only few works have, for the time being, focused on \btledbat\  aspects~\cite{cohen10iptps,pam10,icccn10,globecom10,lcn10,gordon2010iccnt}.

An experimental methodology is followed in \cite{cohen10iptps,pam10}.
In ~\cite{cohen10iptps}, BitTorrent developers detail a specific aspect of their
implementation: namely, an algorithm to solve the problem of the clock drift in \btledbat, to ameliorate the queuing delay estimation at the sender side.
In \cite{pam10}, we present an experimental study of the protocol, exploiting a black-box approach, since at the time of the experiments the protocol was closed-source.

Most of the work on \btledbat\ however adopts a simulative approach~\cite{icccn10,globecom10,lcn10,gordon2010iccnt}. In our previous work~\cite{icccn10}, we performed a preliminary performance evaluation of \btledbat\, considering the default parameter settings suggested in the IETF draft, and unveiling the latecomer issue. In \cite{lcn10}, we instead focus on a comparison of low-priority protocols, contrasting \btledbat\ with TCP-LP~\cite{tcp_lp} and TCP-NICE~\cite{tcp_nice} and performing a sensitivity analysis of the protocol parameters. Along similar lines, authors in~\cite{gordon2010iccnt} investigate the policies for dynamic parameter tuning.
In~\cite{globecom10} we instead focus closely on the fairness issue, and  identify the late-comer advantage as an intrinsic drawback of Additive Increase Additive Decrease (AIAD) policy, as was already shown by Jain in the 80s~\cite{Jain80}. In the case of \btledbat\, we show in \cite{globecom10} that errors in the queuing delay measurement can further exacerbate the problem.

This work is build upon the knowledge gained in our previous effort~\cite{icccn10,globecom10,lcn10}, from which it significantly differs. First, the methodology adopted in this work is more mature and diversified, as we use a fluid approach and analytical techniques as well as simulations, which was the sole technique adopted in \cite{icccn10,globecom10,lcn10}. Also, the scenarios investigated are more complex than our previous work, providing a more complete picture of the protocol performance. Moreover, with the exception of \cite{globecom10}, our previous work mainly focused on the evaluation of the \btledbat\ as is, i.e., without attempting any modification to the draft proposal. Finally, even though we proposed some simple solutions to the fairness issue in \cite{globecom10}, these partly failed in meeting the efficiency goal as well -- which we instead successfully address in this work.

\section{Motivation}\label{sec:motivation}

\begin{figure}[t]
\centering
        \includegraphics[angle=-90,width=0.45\textwidth]{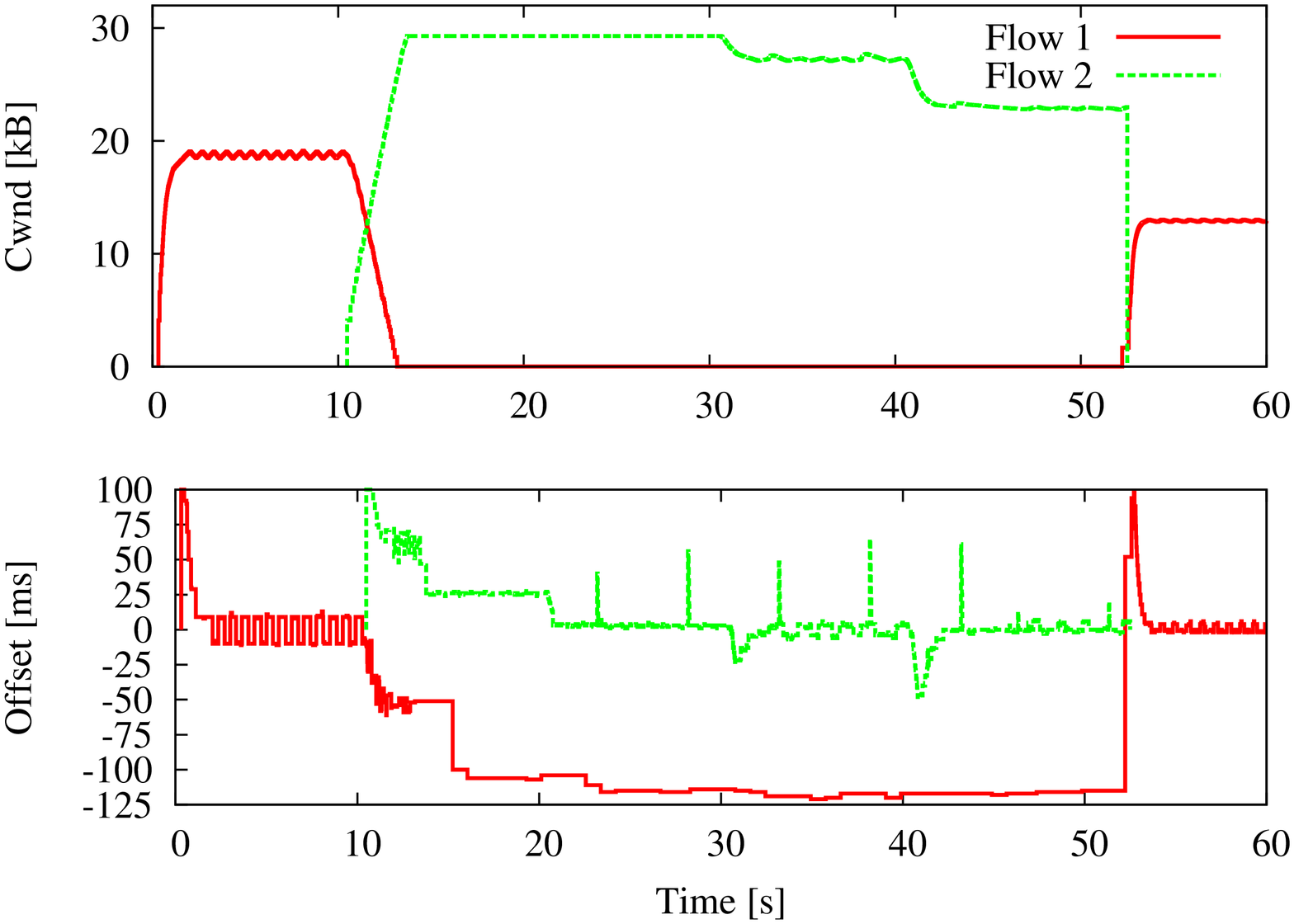}
	\figLC{motivation}{Experimental LAN testbed: Congestion window evolution (top) and offset from the target (bottom) for two competing libUTP flows.}
\end{figure}

Recently, BitTorrent has released an open-source \btledbat\ library~\cite{libutp_code} named libUTP, that we use to show that the latecomer unfairness unveiled in \cite{icccn10} by simulation, also hold in practice. As in \cite{icccn10}, we consider two PCs connected by a $C=10$\,Mbps Ethernet bottleneck, where we emulate by means of \texttt{netem}~\cite{netem} a $RTT=50$\,ms delay.
The first flow starts at time $t=0$ while we let the latecomer join (and spoil) the party at $t=10$\,s. Backlogged transfers are started using the source code provided in~\cite{libutp_code}, instrumented to produce detailed application-level logs\footnote{Packet level traces are also captured and post-processed as in \cite{pam10} for cross-checking purposes: the results, which we are unable to report here for lack of space, are in agreement with the application logs.}.
Results of the experiment are shown in \figR{motivation}, whose top portion reports the time evolution of the congestion window of the two flows. As soon as the first flow starts, it increases its congestion window until the target is reached, and then settles. However, when the latecomer kicks in at $t=10$\,s, the congestion window of the first-comer drops until starvation. The situation persists until $t=50$\,s, time at which we stop the latecomer transfer: right after, the first-comer opens its congestion window again, saturating the link.

This behavior can be explained considering that \btledbat\ aims at introducing a \emph{fixed target} amount of queuing delay in the bottleneck.
The bottom plot of \figR{motivation} reports the time evolution of the offset from the fixed queuing delay target\footnote{Notice that while the IETF draft specifies a mandatory value of $\tau=25$\,ms, the BEP29 document and the implementation actually use $\tau=100$\,ms, which, as shown in \cite{lcn10}, can be cause of further unfairness.} \mbox{$\tau=100$\,ms} measured by each \btledbat\ flow. At $t=0$ the queue is empty, so the queuing delay is null and the offset sensed by the first flow equals the target. As the first flow grows its window and starts transmitting causing queuing delay, the offset shrinks, until the target is hit and the offset is reaches zero: in this region, the congestion window settles and the capacity is efficiently exploited. However as soon as the second flow starts, it senses a non-null queuing delay: more precisely, it senses a queuing delay equal to the target, which is caused by the first comer, to which it adds its own target $\tau=100$\,ms. The latecomer thus sets a target higher than the first one (namely, double in this case),  and aggressively starts climbing the bottleneck. In its phase, the first comer senses a growing queuing delay, which exceeds its own target (negative offset from target means target exceeded) and so it slows down its own sending rate. This is an unfortunate situation, that can however be easily corrected as we show in the following sections.

\section{Proposed LEDBAT modification}\label{sec:model}

 According to the original draft proposal\cite{ledbat_draft}, \btledbat\  maintains a minimum one-way delay estimation $D_{min}$, which is used as base delay to infer the amount of delay due to queuing. \btledbat\ flows have a target queuing delay $\tau$, i.e., they aim at introducing a small, fixed, amount of delay in the queue of the bottleneck buffer. Flows monitor the variations of the queuing delay $q(t) - D_{min}$ to evaluate the distance $\Delta(t)$ from the target:
\begin{align}\label{eq:ledbat_delta}
   &	\Delta(t) = (q(t) - D_{min})-\tau,
\end{align}
where  $q(t)$ is the queueing delay measured at time $t$.
The value of the offset $\Delta(t)$ is then used to drive the congestion window evolution, which is updated packet-by-packet at each acknowledgement reception as it follows:
\begin{align}\label{eq:cwnd_packet}
   &cwnd(t+1)=
   \begin{cases}
      cwnd(t)+\alpha \frac{\tau - \Delta(t)}{\tau} \frac{1}{cwnd(t)}&\text{if no loss },\\
      \frac{1}{2}cwnd(t) &\text{if loss}.
   \end{cases}
\end{align}
where $t$ is a discrete time variable that increments by 1 at each ack arrival and $cwnd(t)$ is the congestion window at time $t$.
The drawbacks of such a congestion window update mechanism have been outlined in \cite{globecom10} and mainly 
consist in the intra-protocol unfairness coupled with a poor calibration of the \btledbat\ level of (low) priority 
w.r.t standard TCP.

We proved in \cite{globecom10} that the unfairness rising among two competing \btledbat\ flows starting at different moments is due to the
\emph{additive decrease} component $\alpha \frac{\tau - \Delta(t)}{\tau}$  that intervenes when $\Delta(t)>\tau$.
We therefore propose to modify the delay-based decrease term and \emph{to introduce a multiplicative decrease} continuously driven by the estimated distance from the target, $\Delta(t)$.
Clearly, to guarantee at the same time fairness and protocol efficiency, a proper choice of the decrease factor has to be made as we will observe in the following sections, as to prevent significant drops in the congestion window.
In addition, we observe that the additive increase term as in (\ref{eq:cwnd_packet}) leads \btledbat\ flows to slow down the increase factor until the target $\tau$ is reached, in which case the window increase completely stops.
This clearly implies a smaller convergence to the target and hence a minor efficiency if compared to the case of a constant additive increase factor  independent of $\Delta(t)$. Based on the above observation, we propose to modify the increase term as well, and \emph{to introduce an additive increase} according to a constant factor $\alpha$ as in TCP Reno.
Notice that in this way, we expect to achieve better efficiency performance without violating the low priority requirements as expressed in the \btledbat\ draft. Indeed, by selecting $\alpha\le 1$ the additive increase component can be made at most as aggressive as TCP.


Summarizing, we propose to modify the congestion window evolution as follows:\\

\noindent $cwnd(t+1)=$
\begin{align}\label{eq:new_cwnd_packet}
   \begin{cases}
      cwnd(t)+\alpha  \frac{1}{cwnd(t)} & \text{if no loss and } \Delta \le0,\\
      cwnd(t)+\alpha  \frac{1}{cwnd(t)}- \frac{\zeta}{\tau} \Delta  & \text{if no loss and } \Delta > 0,\\
      \frac{1}{2}cwnd(t) & \text{if loss}.
   \end{cases}
\end{align}

In the following sections we quantify the overall improvement deriving by such a congestion window update by means of both a \emph{fluid model}, which provides a closed-form characterization of the stationary throughput and \emph{simulations}, which allow the study of more complex scenarios.
In the remainder of this paper, we refer to the modified version of \btledbat\ as fair-\btledbat\ (\fledbat).

\subsection{Fluid Model description}\label{sec:fluidmodel}
\begin{table}[!t]
   \begin{footnotesize}
   \begin{center}
      \caption{Notation}\label{tab:notation}
      \begin{tabular}{|l|l|}
         \hline
          $N$                      & Number of \fledbat\ flows \\
          $C$                      & Link capacity \\
          $\{W^i(t)\}_{i=1,...,N}$ & Congestion windows at time t \\
          $\{X^i(t)\}_{i=1,...,N}$ & Instantaneous rates at time t \\
          $Q_t$                    & Queue occupancy at time t \\
          $\alpha$                 & Additive Increase factor \\
          $\zeta$                  & Multiplicative Decrease factor \\
          $R_t$                    & Round trip time at time t \\
          $\tau$                   & Queuing delay target \\
         \hline
      \end{tabular}
   \end{center}
   \end{footnotesize}
\end{table}
In this section we develop a fluid model of the congestion window and hence of the transmission rate of one or more \fledbat\ flows aimed at capturing first order system dynamics.
The congestion window is now a continuous variable both in time and in space, $W(t)$  (the notation is summarized in Tab.\ref{tab:notation}).
We consider the case of $N$ \fledbat\ flows sharing the same link of capacity $C$ and experiencing the same\footnote{Though the model generalizes to the case of heterogeneous RTT, for the sake of simplicity in this paper we focus on the homogeneous case.} Round Trip Time $R_t$.
In addition, we make the following assumptions:
\begin{itemize}
  \item The round trip time $R_t$ is defined by the sum of twice the propagation delay, $R$, transmission delay $1/C$ and queueing delay $q(t)$.
  We further assume that the propagation delay is predominant, i.e. $R_t\approx R$.
   \item The queueing delay $q(t)$ is defined as ratio of the queue occupancy $Q_t$ at time $t$ divided by the link capacity $C$, i.e.,   $q(t)=Q(t)/C$. Thus, we assume that the queuing delay information \emph{instantaneously} propagates to the sender, neglecting thus the delay in the feedback loop.
\item We further assume that flows can correctly estimate the queuing delay, which is equivalent to take $D_{min}=0$.
  \item By Little's law, we assume that congestion windows and link rates are linked by:
\begin{align}
   &X^i_t= W^i_t/R_t, \quad \forall i=1,...,N
\end{align}
\end{itemize}

Remark that the assumption that flows can correctly estimate the queuing delay may again not hold in practice.
As such, we expect that simulation results may show an offset with respect to the model predictions, which is due to such simplifying assumption.
There are however two main reasons for which we believe these assumption, that make the problem tractable, are also reasonable.
On the one hand, additional mechanisms to enhance the delay estimation accuracy could be then adopted in order to ameliorate the overall protocol performance (as it has been done in previous work \cite{tcp_lp}, and which is indeed part of the current BitTorrent effort~\cite{cohen10iptps}, hence reducing the error and reinforcing our assumptions). On the other hand, a more fundamental reason is that the characterization of protocol dynamics in absence of such estimation error is a necessary step in the \fledbat\ protocol design -- as, even though on simplistic settings,  important properties of the protocol such as efficiency and fairness can be \emph{proved} to hold with the help of a rigorous framework.

\subsection{Fluid system dynamics}
Let us consider the case of a single \fledbat\ connection, whose congestion window evolves according to (\ref{eq:new_cwnd_packet}).
The corresponding flow-level congestion window evolution is:
\begin{align}\label{eq:W}
   \frac{dW(t)}{dt}= \frac{\alpha}{R}-\frac{\zeta}{\tau} \left(\frac{Q(t)}{C}-\tau \right)\frac{W(t)}{R}
\I{W(t)\ge } \I{Q(t)\ge C\tau},
\end{align}
where we denote by $W(t)$ the instantaneous congestion window at time $t$ in the fluid system. As we assume an approximately constant round trip delay, we
replace $R_t$ by $R$ in (\ref{eq:W}).
The instantaneous queue occupancy instead satisfies:
\begin{align}\label{eq:Q}
   \frac{dQ(t)}{dt}=\frac{W(t)}{R}-C \I{Q(t)\ge 0}.
\end{align}

\noindent where, in other words, only the flow that exceeds the capacity creates queuing in the buffer.
Thus, the instantaneous rate, $X(t)$, satisfies:
\begin{align}\label{eq:X}
   \frac{dX(t)}{dt}= \frac{\alpha}{R^2}-\frac{\zeta}{R\tau}\left(\frac{Q(t)}{C}-\tau \right) X(t)\I{X(t)\ge } \I{Q(t)\ge C\tau}
\end{align}
and (\ref{eq:Q}) can be re-written as
\begin{align}\label{eq:Q2}
   \frac{dQ(t)}{dt}=X(t)-C \I{Q(t)\ge 0}.
\end{align}

\subsection{Main results}
We now present the main results of this paper: namely, the existence of a unique and globally stable solution. 
We also express, with closed form formul\ae, the performance of the protocol at the equilibrium, proving its 
\emph{efficiency} and \emph{fairness} -- which was our initial goal. Let us start by proving that the system admits 
a unique solution.
\begin{prop}\label{prop:one-flow1}
The system of ODEs (\ref{eq:X})-(\ref{eq:Q2}) admits the unique
equilibrium $P^*=(X_1^*,\dots, X_N^*,Q^*)$
\begin{align}\label{eq:steady-XQ}
  &X_i^*=C/N,~i=1,\dots,N  &Q^*=C\tau+\frac{N\alpha\tau }{\zeta R}
\end{align}
where $X_i^*$ and $Q^*$ denotes the stationary values of $X_i$ and $Q$ respectively.
\end{prop}
\vspace{-1mm}
\begin{proof}
Be $X=\sum_i^N X_i$, we consider the stationary regime by the condition
$(\dot{X}_i,\dots, \dot{X}_N,\dot{Q})=(0,\dots,0)$
\begin{align}
\dot{Q} &= 0 \Leftrightarrow X^*=C, &\nonumber \\
\dot{X_i} &= 0 \Leftrightarrow 0=\frac{\alpha}{R^2} - \frac{\zeta}{R C\tau } (Q^*-C\tau)X^*,\nonumber \\
&\Leftrightarrow 0=\frac{\alpha}{R^2}-\frac{N\alpha}{C R^2}X^*_i\Leftrightarrow X_i^*=C/N,~i=1,\dots,N.
\end{align}
\end{proof}
\noindent Then, the following proposition states that this unique equilibrium
is also globally stable (see \cite{Verhulst}).
\begin{prop}\label{prop:one-flow2}
The system of ODEs (\ref{eq:X})-(\ref{eq:Q2}) is globally stable in $P^*$.
\end{prop}

\begin{IEEEproof}
We consider the trajectories of the point $(X,Q)\in\mathbb{R}^2$ driven
by the ODEs (\ref{eq:X})-(\ref{eq:Q2}).
In the region $A=\{x,q : 0<q<C\tau\}$ the state equations simplifies to
\begin{align}
\begin{cases}
\dot{X}  &= \frac{N\alpha}{R^2}\Rightarrow X_t = X_0+\frac{N\alpha}{R^2}t\\
\dot{Q}  &= X-C\Rightarrow Q_t=Q_0+(X_0-C)t+\frac{N\alpha}{2R^2}t^2
\end{cases}
\end{align}
Clearly, for any $(X_0,Q_0)\in A$  there exists a finite $t\geq 0$ such that
$(X_t,Q_t)\notin A$. This means that all point in $A$ are unstable.  
For $(X,Q)\notin A$  the state equations becomes
\begin{align}
\begin{cases}
\dot{X}  &= \frac{N\alpha}{R^2}-\frac{\zeta}{C R\tau}(Q-C\tau)X \nonumber\\
\dot{Q}  &= X-C
\end{cases}
\end{align}
Outside $A$ we define the Lyapunov function, 
\begin{align} 
V(X,Q)=(X-X^*)-\log\left(\frac{X}{X^*}\right)+\frac{\zeta (Q-Q^*)^2}{2R C\tau}
\end{align} 
Clearly, $V(P^*)=0$, $V(X,Q)\geq 0$ $\forall (X,Q)\in \mathbb{R}^2$.
\begin{align} 
\dot{V}(X,Q)
&=\dot{X}-\dot{X}\frac{X^*}{X}+\dot{Q}\frac{\zeta(Q-Q^*)}{R C\tau}\nonumber \\
&=\frac{\dot{X}}{X}(X-X^*)+(X-X^*)\left[ \frac{\zeta(Q-Q^*)}{R C\tau}\right]\nonumber \\
&=(X-X^*)\left[\frac{N\alpha}{X R^2}-\frac{\zeta(Q-C\tau)}{RC\tau}+\frac{ \zeta(Q-Q^*)}{RC\tau}\right]\nonumber \\
&=(X-X^*)\left[\frac{N\alpha}{X R^2}-\frac{N\alpha}{C R^2}\right]\nonumber \\
&=-\frac{N\alpha}{X R^2}(X-X^*)^2 <0
\end{align} 
Which states that $P^*$ is an equilibrium globally stable.
\end{IEEEproof}

\begin{prop}\label{prop:one-flow3}
The system of ODEs (\ref{eq:X})-(\ref{eq:Q2}) is locally
stable in the equilibrium $P^*=(X_1^*,\dots, X_N^*,Q^*)$
\end{prop}
\begin{IEEEproof}
We write $(\dot{X}_1,\dots,\dot{X}_N,\dot{Q}) = (f_1,\dots,f_N,g)$, for $X_i>0$, $Q>0$ where $f_i$ and $g$
are defined as follows:
\begin{align}\label{eq:local-stable-XQ}
\begin{cases}
f_i(X,Q) =& \frac{\alpha}{R^2} - \frac{\zeta}{C\tau R}(Q-C\tau)X_i\I{Q(t)\ge C\tau} i=1,...,N\\
g(X,Q) =& X - C
\end{cases}
\end{align}
Linearizing the system of ODEs in $P^{*}$, and defining  $\Delta X_i = X_i-X_i^{*}$, $\Delta Q = Q-Q^{*}$,
and $\mathbf{Y}=(\Delta X_1,\dots,\Delta X_N,\Delta Q)$ we obtain $(\tilde{f}_1,\dots, \tilde{f}_N,\tilde{g})= \dot{\mathbf{Y}}=  A \mathbf{Y}$ where $A$ is a $(N+1)\times (N+1)$ square real matrix defined as follows:
$$A= \left(
\begin{array}{ccccc}
-\frac{\alpha}{C R^2} &  0 & \cdots & 0 & -\frac{\zeta}{C\tau R } \\
	0 	      &  -\frac{\alpha}{C R^2} & \cdots & 0 & -\frac{\zeta}{C\tau R } \\
       1              &  1 & \cdots & 1 & 0
\end{array}
\right)$$
The characteristic polynomial is then
$\left(\lambda + \frac{\alpha}{C R^2}\right)^{N-1}\left(\lambda^2+\frac{\alpha}{C R^2}\lambda +N\frac{\zeta}{C\tau R }\right)$
whose roots have all real part negative.
\end{IEEEproof}
\begin{prop}\label{prop:one-flow4}
The solution of the system of ODEs (\ref{eq:X})-(\ref{eq:Q2}) converges to the
global stable equilibrium $P^*$ at a rate $e^{-\Theta t}$ with,
$\Theta=\frac{\alpha}{CR^2}\left(\frac{1+\I{\zeta\leq \zeta^*}\sqrt{1-\zeta/\zeta^*}}{2}\right)$
and $\zeta^*=\frac{\alpha^2 \tau}{4 N C R^3}$.
\end{prop}
\begin{IEEEproof}
We calculate the dominant eigenvalue of
the matrix $A$, i.e. the eigenvalue with the real part
with the smallest absolute value.
\end{IEEEproof}
\noindent To conclude, we summarize our main findings in the following observation, expressing the results in terms 
of the expected performance of \fledbat.
\begin{obs}
Prop. \ref{prop:one-flow1},\ref{prop:one-flow2},\ref{prop:one-flow3},\ref{prop:one-flow4}
prove that the designed protocol is efficient ($X^*=C$), and long term fair
($X_i^*=C/N$). In addition the queuing delay attains the target $\tau$
($Q^*/C = \tau + \frac{N\alpha\tau }{ C \zeta R}$) by an error of $\frac{N\alpha\tau }{ C \zeta R}$.
\end{obs}

\noindent Thus, our initial goals of an \emph{efficient} and \emph{fair}
protocol is met. Clearly, a  number of issues need further investigation  (i.e., how the protocol performs in practice where not all modeling assumption holds, what is the impact of parameters and of packet-level dynamics, how does it performs against TCP, etc.) that we dig in the next section by means of a thorough simulation campaign in a number of different scenarios.

However, prior to conclude this section, we still need to provide evidence of the accuracy of the model, which we do by comparing the numerical solution of the fluid model with results gathered with packet-level \texttt{ns2}~\cite{ns2} simulations (our implementation is available as open source code at \cite{ledbat_code}). As far as the network is concerned, we consider a simple scenario with a  $C=10$\,Mbps bottleneck, $RTT=50$\,ms and buffer size $B=100$ packets. As far as the application is concerned, we consider two \fledbat\ flows sending $P=1500$\,Byte full-payload packets, with the same target $\tau=25$\,ms. For the time being, we fix the decrease component by setting $\zeta=0.1$, and explore the impact of $\zeta$ later on. To recreate the conditions for the latecomer unfairness phenomenon,
the two flows do not start at the same time, but their start time are separated by $2$\,seconds. The time evolution of the system state is in  \figR{comparison}, which reports both the evolution of the flow rates $\{X_i\}$ (top) and the buffer occupancy $Q_t$ (bottom) gathered either by numerical solution (right) or \texttt{ns2} simulation (left).  As a general comment, the numerical solution is in agreement with the simulation results, although the packet-level dynamic exhibits much wider fluctuations, which is expected, since fluid model gives an average behavior. Moreover, on average the fluid system dynamics closely matches the results gathered via simulation (for the sake of readability, we plot the moving average of the queue length gathered via simulation alongside the instantaneous occupancy).  As expected, both numerical and simulation results show that the capacity is, after an initial transient phase, fairly shared among flows (i.e., $\overline{X}_i\approx C/2, \forall i$) and that furthermore the queuing delay target is reached (i.e., $\overline{Q}_t \approx C \tau$).

\begin{figure}[t]
    \begin{center}
       \subfigure[]{ \includegraphics[angle=-90,width=0.45\textwidth]{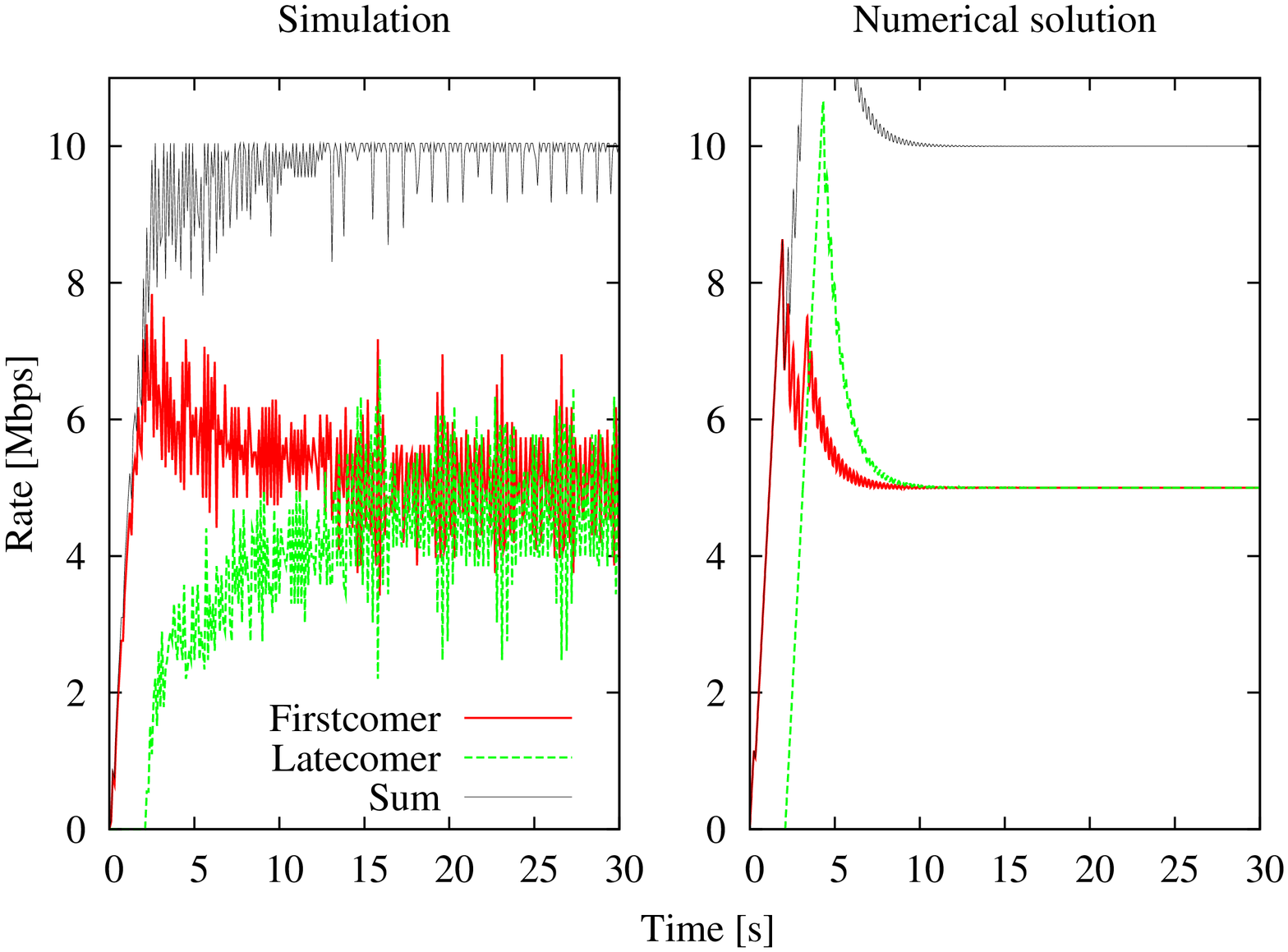}}
       \subfigure[]{ \includegraphics[angle=-90,width=0.45\textwidth]{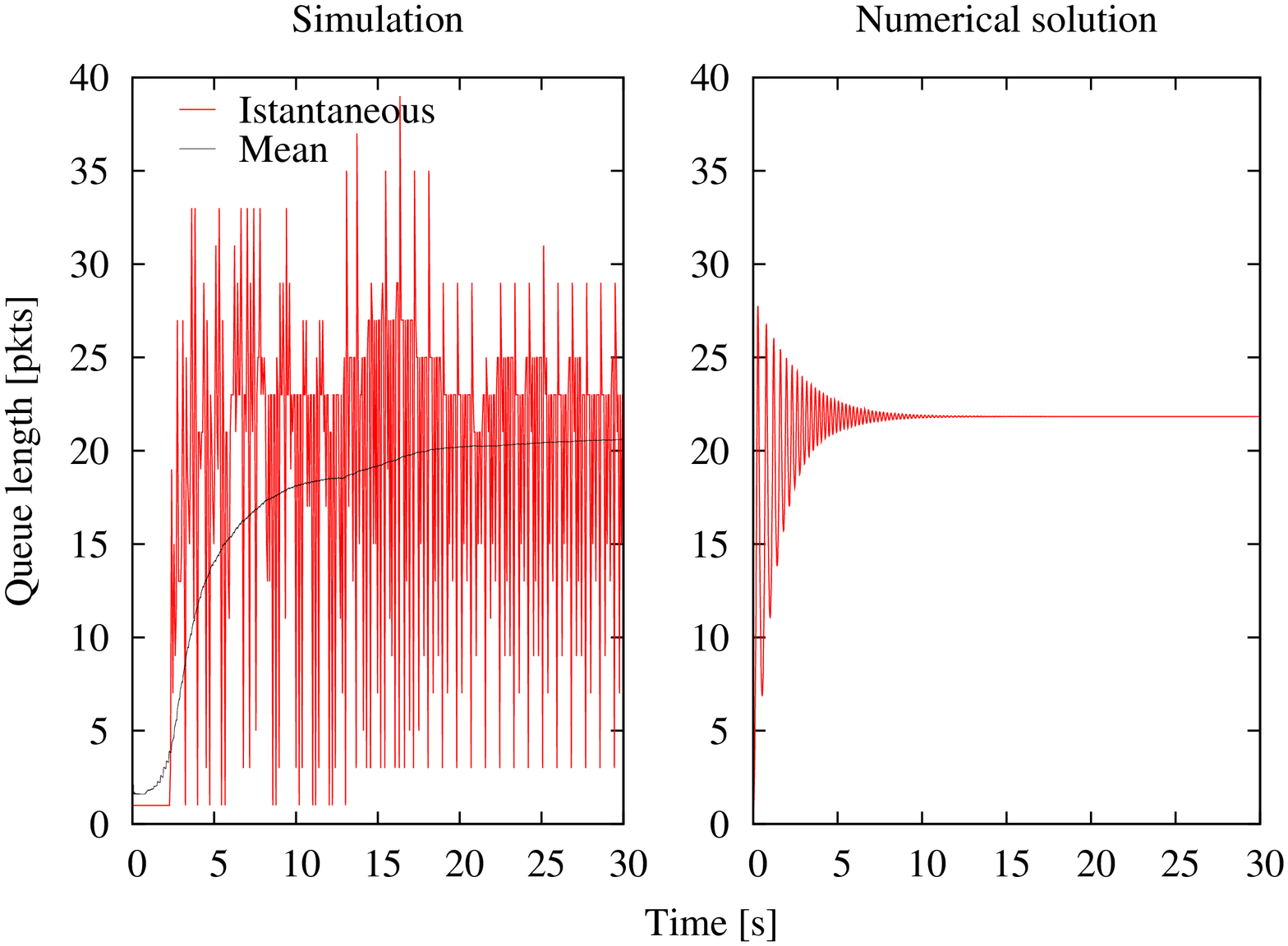}}
        \figLC{comparison}{Comparison of (left) simulation and (right) numerical solution for (a) Rates and (b) Queue length}
    \end{center}
\end{figure}

\section{Simulation results}\label{sec:simulation}
%
%
\begin{figure*}[t]
    \begin{center}
       \subfigure[]{ \includegraphics[angle=-90,width=0.45\textwidth]{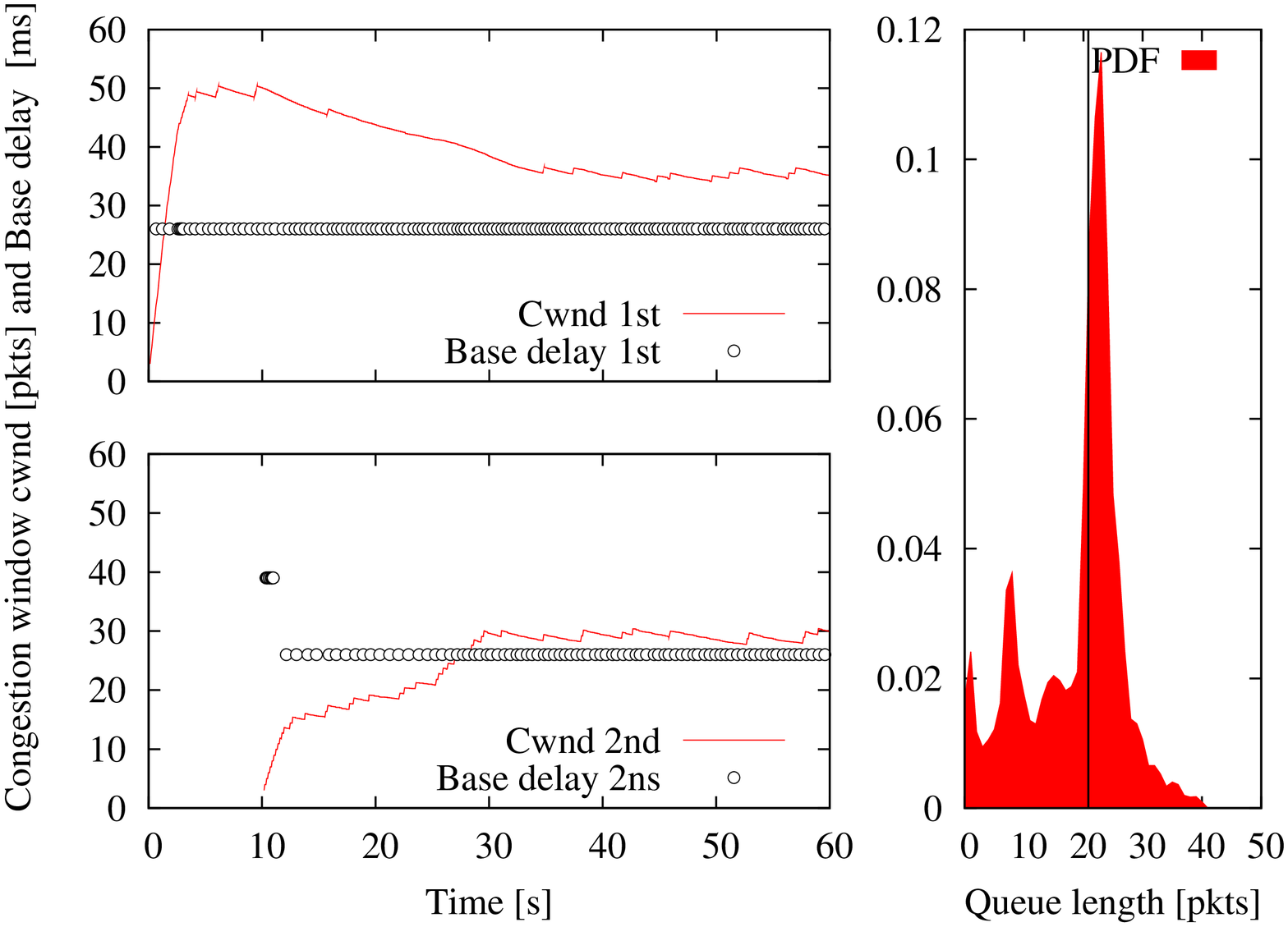}
       }
       \subfigure[]{ \includegraphics[angle=-90,width=0.45\textwidth]{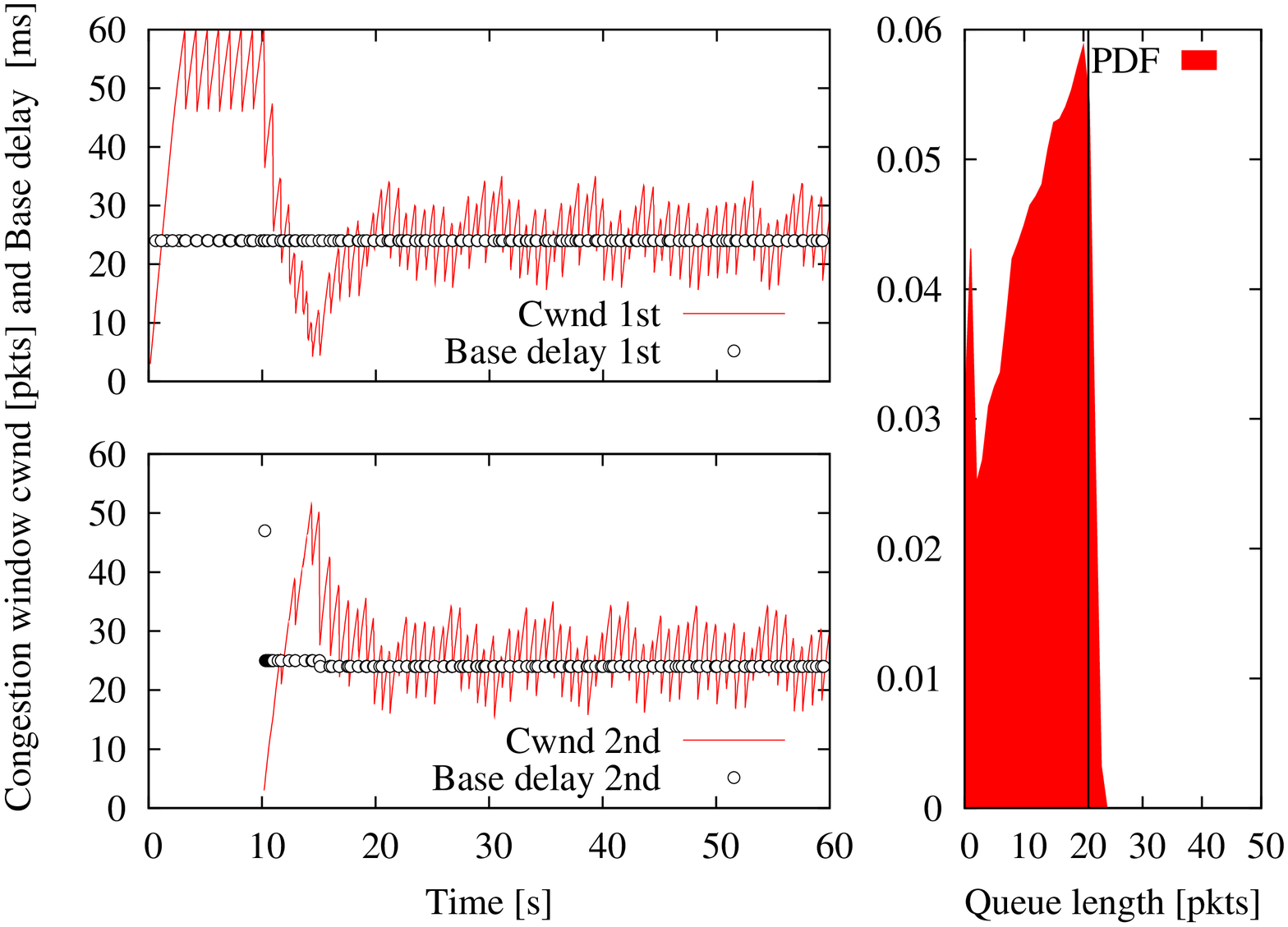}
}
        \figLC{time}{Time evolution of \fledbat\ dynamics with (a) chunk-based $\zeta=0.01$ and (b) backlogged $\zeta=5$ traffic models.}
    \end{center}
\vspace{-4mm}
\end{figure*}

In the previous section we have, first, developed a mathematical model of the
new protocol in order to formally prove its properties and, second, shown that the fluid model dynamics correctly matches the protocol behavior we observe through packet level simulations in a simple scenario.
However, as previously stated, the model is based on a number of simplifying assumptions
 and it furthermore neglects some aspects due to packet-level quantization (i.e., queue length and congestion window in multiple of fixed-size packets as opposite to continuous rate in the fluid model). Yet, we show that our protocol design is robust even in presence of such simplifying assumption by carrying on a simulative study of the \fledbat\ protocol.

Unless otherwise stated, we consider the reference scenario of \cite{icccn10}, consisting in a bottleneck link of capacity $C=10$\,Mbps and buffer size $B=100$ packets. For the sake of simplicity we consider fixed size packets of $P=1500$\,Bytes. Data flows in a single direction, and acks are not delayed, dropped nor affected by cross-traffic on their return path.  All flows have the same round trip time $RTT=50$\,ms, half of which is due to the propagation and transmission delay components of the bottleneck link (i.e., a one-way  base delay of 25\,ms). As far as TCP flows are concerned, we select the Reno flavor: in this way we gather conservative results since  we expect more recent TCP variants implemented by default in Linux (Cubic) and Windows (Compound) operating systems to be more aggressive than traditional Reno flows.
Each simulation point reported in the following is the results of $10$ simulation runs, over which we gather the average and standard deviation of the metrics of interest.

In the reminder of this section, we study: (i) the impact of different traffic models, comparing backlogged and chunk-by-chunk transfer; (ii) the sensitivity of the protocol to $\zeta$ parameter variation and (iii) the protocol performance under more realistic and heterogeneous scenarios.

\subsection{Traffic model}

\subsubsection{Chunk-by-chunk transfer}

In this scenario, we consider sources that continuously transmit chunks of
data, where each chunk has the typical BitTorrent size of $250$\,kB (nearly $170$ full payload packets). As soon as a chunk transmission ends (i.e., when the last acknowledgment for that chunk has been received at the sender side), a new chunk transmission is scheduled with the same peer. Notice that this traffic model, which emulates the dynamics of P2P traffic exchange, differs from backlogged transfers in that,
after the last data packet of a chunk has been sent, the source peer stops transmitting for about $RTT$ seconds until the matching acknowledgement is received, and a new chunk transmission can start. Notice also that we keep the congestion window parameter
across chunks (i.e., congestion window is \emph{not} reset between subsequent chunks exchanged with the same peer).

\figR{time}-(a) reports the time evolution of the system dynamics when $\zeta=0.01$: in the left portion, congestion window and base delay estimation of the firstcomer (top) and latecomer (bottom) flows are reported, while the right portion shows the  distribution of the queue length. In this case, it can be seen that, despite the latecomer initially has an incorrect view of the base delay (as in \btledbat), the multiplicative decrease phase of the firstcomer allow the latter to correct its estimate, after which the performance share converges to an equitable state. Due to (i) the continuous adjustment of AI and MD dynamics and (ii) the fact that chunk transmission seldom pauses the transmission, the queue is no longer stable as for the standard \btledbat\ case~\cite{icccn10}, but fluctuates around the occupancy value predicted by the model (represented by a solid vertical line).

%
%
\begin{figure*}[t]
    \begin{center}
       \subfigure[]{\label{fig:sensTCP} \includegraphics[angle=-90,width=0.3\textwidth]{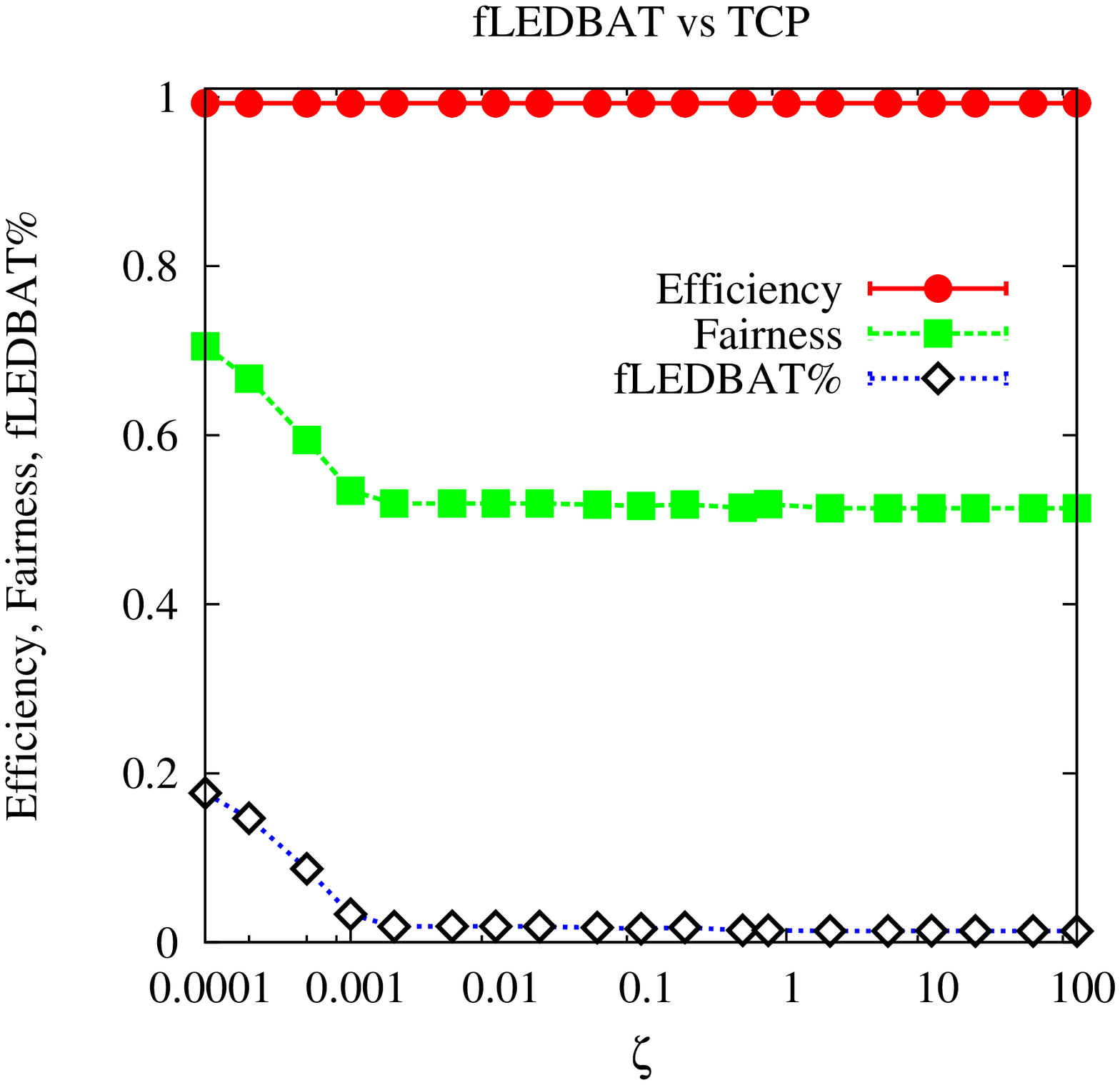}}
       \subfigure[]{\label{fig:sensLEDBAT} \includegraphics[angle=-90,width=0.3\textwidth]{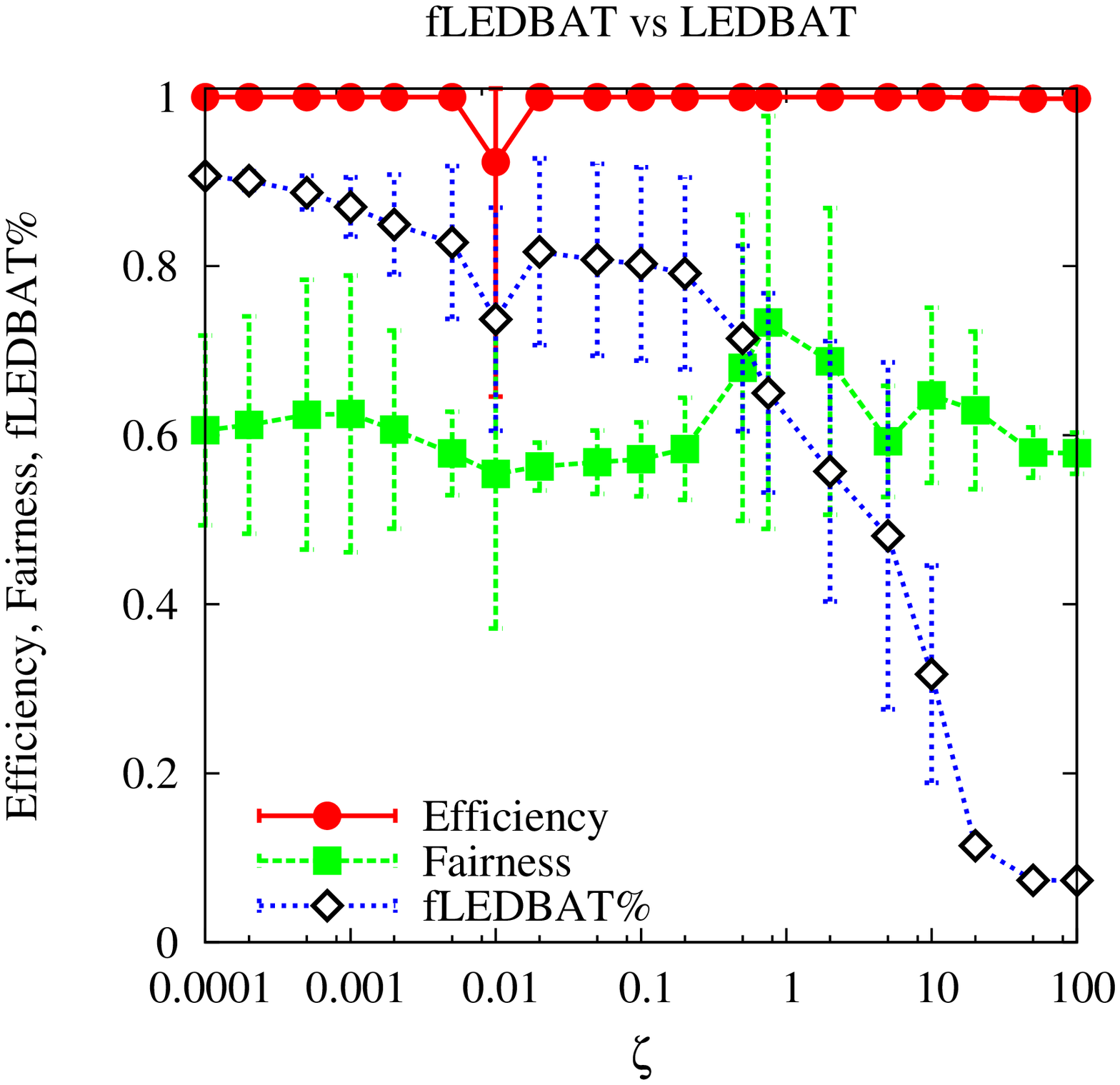}}
       \subfigure[]{\label{fig:sensfLEDBAT} \includegraphics[angle=-90,width=0.3\textwidth]{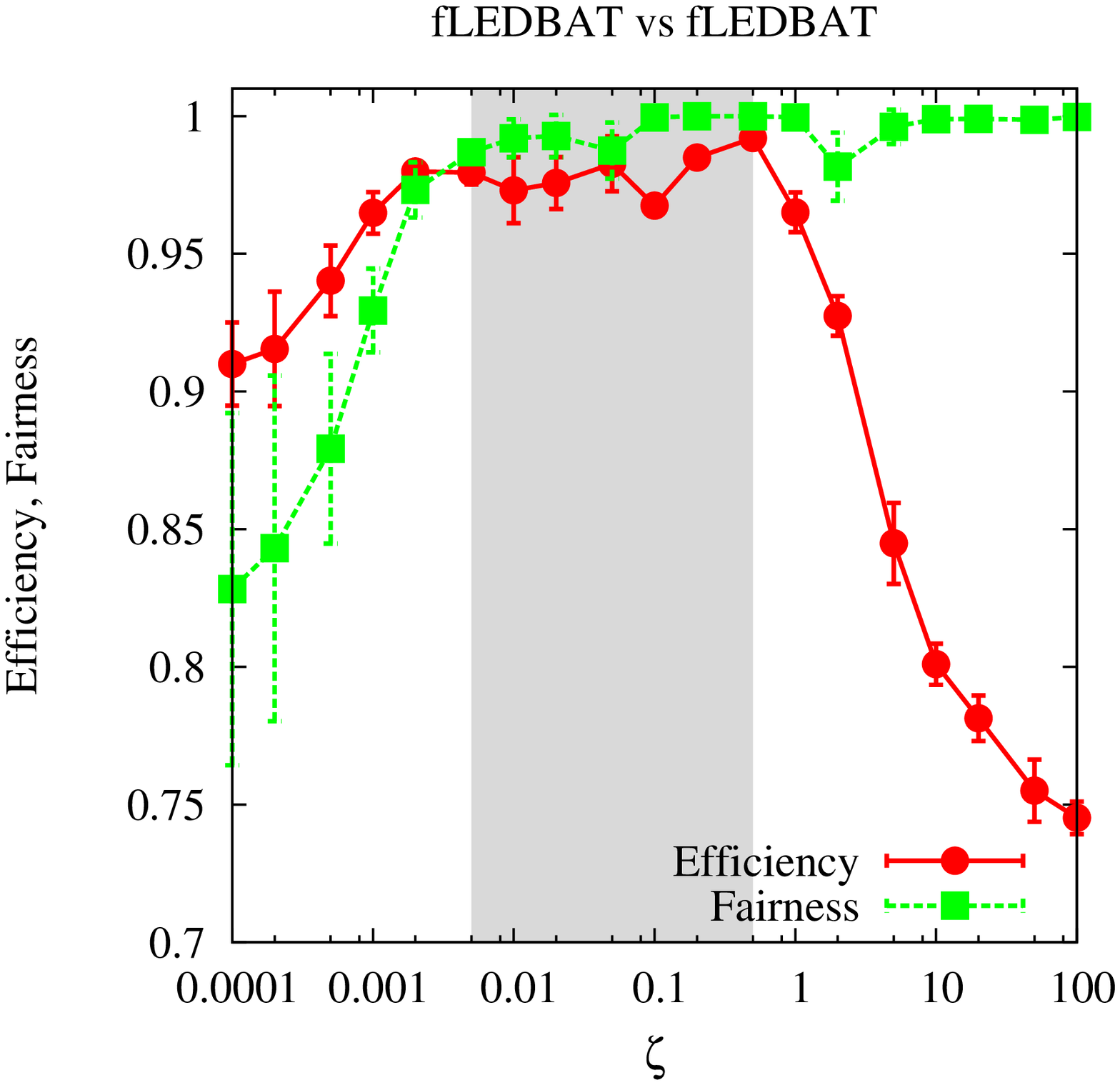}}
       \figLC{sens}{Sensitivity analysis to $\zeta$: Efficiency, long-term fairness and protocol breakdown of a \fledbat\ flow sharing the bottleneck with (a) a TCP flow, (b) a \btledbat\ flow (c) another \fledbat\ flow.}
    \end{center}
\vspace{-4mm}
\end{figure*}

\subsubsection{Backlogged transfer}

In the case of backlogged transmission, a latecomer phenomenon may still arise depending on the value of $\zeta$: indeed, when $\zeta$ is too small, the multiplicative decrease component of the first flow is slower than the additive increase of the latecomer, which is thus unable to correct its wrong estimation.
However, provided that $\zeta$ is large enough to let the queue flush, \fledbat\ can still reintroduce fairness.

Results for the backlogged scenario are reported in \figR{time}-(b) for
$\zeta=5$. Especially, the queue now seldom flushes (as it can be seen by the
increased probability to have a null queue length shown by the PDF)  which helps
latecomers gain a correct view of the base delay. In this case, though, the
model slightly overestimate the queue size: indeed, due to larger $\zeta$
values, the congestion window fluctuations are now wider. Also, as the queue
flushes, the protocol is less efficient with respect to the previous cases too,
because the capacity is not fully utilized all the time.

We point out that, since \fledbat\ is designed to be a low-priority protocol,
slight inefficiency should be tolerable. Conversely, in case efficiency, rather
than low-priority, would have been a more important goal, then an alternative
approach is possible, which we already explored in \cite{icccn10}: indeed, a
simple way to drain the queue empty (which allows each sender to gather correct
measures of the base delay) is to use TCP-like slow-start at flow startup. The
downside in this case, is that as slow-start causes losses, which may have
undesirable side effects on interactive traffic (e.g., VoIP, gaming) and is thus
less indicated in our opinion in this context.

\subsection{Sensitivity analysis}\label{sec:sensitivity}

In this section we carry out further simulations, to assess the impact of the
choice of the parameter $\zeta$ on the protocol performance.
In order to gather a complete sensitivity analysis of \fledbat\ parameters,
we consider several scenarios: (i) a \reno\ flow competing with a
\fledbat\ flow, (ii) a \btledbat\ flow competing with a \fledbat\ flow, (iii) two or more \fledbat\ flows competing for the same bottleneck. All flows operate in chunk-by-chunk transmission mode.

As performance metrics, we consider \emph{fairness}, \emph{efficiency} and \emph{protocol breakdown} of
the data transfer. Specifically, we use Jain fairness index $F$, which is
defined as $F =( \sum_{i=1}^N x_i )^2 /( N \cdot \sum_{i=1}^N x_i^2 )$ where $x_i $ is the rate of the $i$-th flow. We have that fairness tops to $F=1$ when bandwidth is perfectly shared among all flows, while it drops to a minimum of $F=1/N$ when one flow monopolizes the bottleneck leaving the others in starvation. Regarding the efficiency, we consider the link utilization metric $\eta$ defined as the ratio of the overall throughput (including headers) over the link capacity $C$, i.e., $\eta=\sum_{i=1}^N x_i / C$. For the sake of illustration, we also consider the \emph{protocol breakdown}, defined as the percentage of traffic sent by \fledbat\ sources over the total traffic, which immediately conveys the level of low priority of \fledbat\ with respect to other protocols insisting on the same bottleneck.

\subsubsection{Observations on $\alpha$ and $\tau$}

As careful sensitivity analysis focused on gain $\alpha$ and target $\tau$ has already been carried out in \cite{lcn10}, 
in the following we briefly summarize the main lessons as far as these two parameters are concerned, while we provide a thorough set of simulation results for the newly introduced parameter, i.e., the decrease factor $\zeta$.

Let us consider the target parameter $\tau$ first.
Already in the homogeneous case of several flow with equal settings, from \cite{lcn10} we gather that the performance of \btledbat\ can not  be easily controlled by tuning the target $\tau$. Indeed, the low priority level can be changed only when the  $C \tau$  product approaches the buffer size -- however changes in the priority level are too steep for very small variations of $\tau$. Moreover, there is no single value of $\tau$ that can adapt to both low-capacity and high-capacity links at the same time.
Finally, in the heterogeneous case of several flows with different settings, even a small difference between values of $\tau$ yield to extremely unfair situations, with flows having larger $\tau$ being most aggressive. For this reason,  we adhere to the mandatory value specified by the draft  $\tau=25$\,ms and do not consider $\tau$ as a free parameter.

Let us now consider the gain parameter $\alpha$: in this case, is worth noting that the increase component of \fledbat\ differs from that of \btledbat. Indeed, \btledbat\ increase is proportional (with $\alpha$ proportionality constant) to the offset from the target, meaning that as the estimated queueing delay approaches the target, the congestion window growth slows down. In the case of \fledbat\ instead, the congestion window growth is still proportional to $\alpha$, but constant (i.e., no longer dependent on the offset from the target). Therefore, the value of $\alpha=1$ is constrained in reason of the low-priority goal (so to match the $1$-packet-per-RTT TCP growth in congestion avoidance).

\subsubsection{\fledbat\ vs TCP}

\figR{sensTCP} shows the efficiency and fairness performance when a single
TCP and a single \fledbat\ flow share the bottleneck: first of all, we can see that low-priority goal is met, as TCP is enjoying the largest portion of the capacity (\fledbat\ breakdown goes to 0\% and fairness drops to $1/N$).
As expected, efficiency is high: as we already observed in \cite{icccn10} for \btledbat, \fledbat\ is still able to push some bytes on the link, thereby increasing the overall link utilization with respect to the case where a single TCP Reno flow insists on the bottleneck.

With the exception of extremely low values of $\zeta<10^{-3}$  (which soften the effect of the multiplicative decrease, and sharpen the impact of the Reno-like additive increase), the low-priority goal is therefore satisfied. Thus, selecting $\zeta$ is not a concern as far as heterogeneous \fledbat\ vs TCP scenarios are considered.

\subsubsection{\fledbat\ vs \btledbat}

\figR{sensLEDBAT} shows the efficiency and fairness performance when a single \fledbat\ flow and a \btledbat\ one share the bottleneck. We randomize the start time of both flows in the [0,10]\,sec interval, so that the latecomer can be either of the two protocols.
In this case, we gather \fledbat\ is generally more aggressive (due to the AI dynamic, which is more aggressive than \btledbat\ PID dynamic) until $\zeta$ grows too large, in which case the reverse happens (due to the MD dynamic being more drastic than the PID dynamic).
Specifically, less than 20\% of the bottleneck is occupied by the \btledbat\ when  $\zeta<10^{-2}$. For larger values of $\zeta$ though, \btledbat\ becomes increasingly competitive with  \fledbat: the crossover happens at about $\zeta=5$, after which \fledbat\ becomes even lower priority than \btledbat.

In no case, however, the share is fair and a latecomer phenomenon may still arise. Consider that, when \btledbat\ start first and saturates the bottleneck, it induces a very steady queue. Therefore, when an \fledbat\  latecomer flow arrives on the bottleneck, it measures an incorrect base delay. However, as \btledbat\ reacts with a \emph{linear} decrease to the increasing delay, the \fledbat\ latecomer will not have the opportunity to correct is estimate -- as it otherwise does whenever the firstcomer flow reacts with a \emph{multiplicative} decrease to the increasing delay. Hence, when $\zeta$ is small, the \fledbat\  latecomer can starve the \btledbat\ flow.

\subsubsection{\fledbat\ vs \fledbat}

We finally consider the intra-protocol scenario in which $N\ge2$ \fledbat\ flows share the bottleneck. We start by considering the $N=2$ case and set the start time of latecomer flow to $t=10$\,s, which was shown in \cite{icccn10} to
represent a worst case scenario for the fairness index.
\figR{sensfLEDBAT} reports results for varying $\zeta$, where we omit this time the \fledbat\ breakdown. From the picture, it is clear  that \fledbat\ is able to operate fairly and efficiently under a wide range of parameters. Overall, taking into account also the previous remark in the intra-protocol \fledbat\ vs TCP scenario, we have that any value of $\zeta$ in the gray shaded zone yield to an efficient, fair  and low-priority system.

%
%
\begin{figure}[t]
    \begin{center}
       \includegraphics[angle=-90,width=0.45\textwidth]{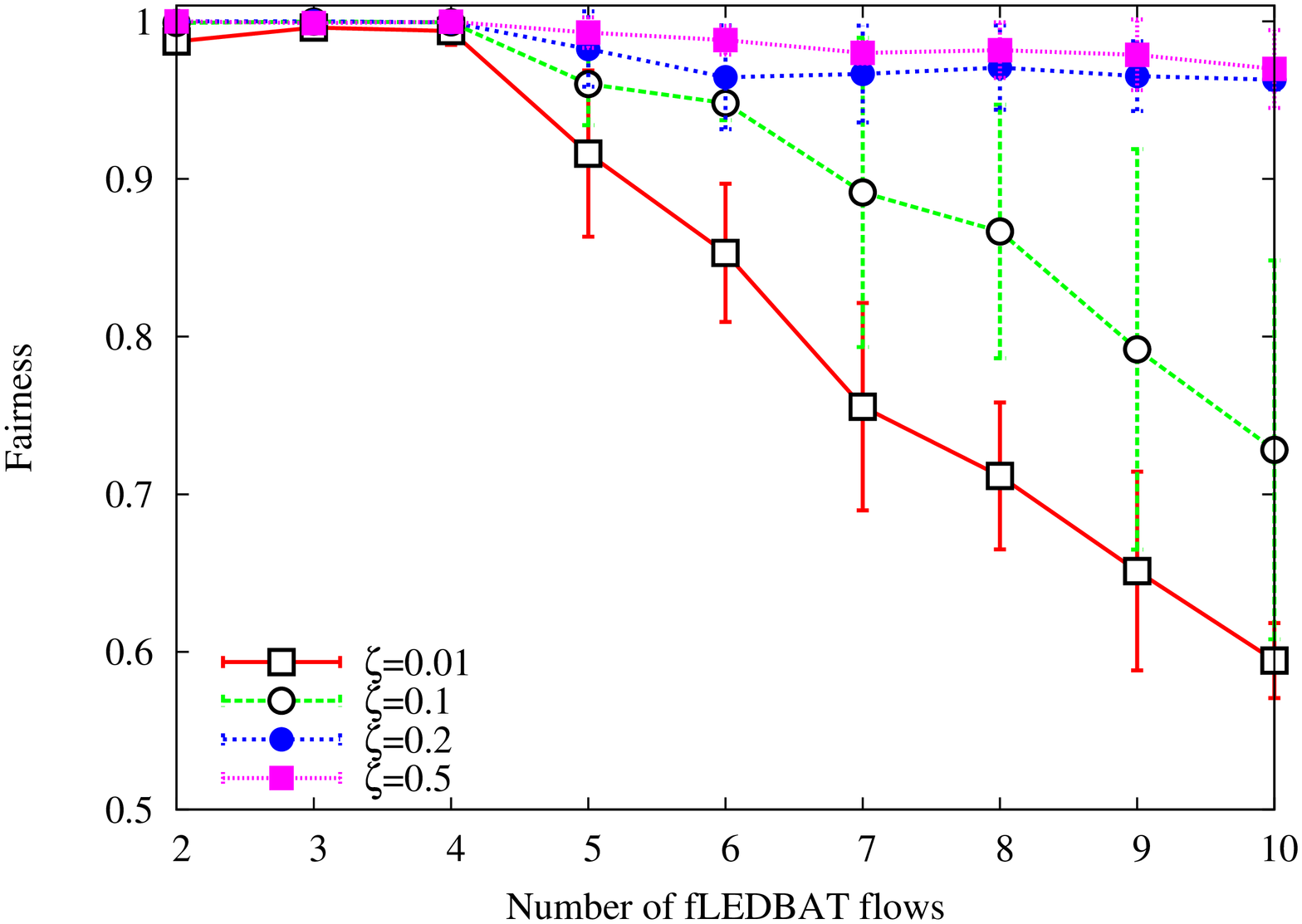}
        \figLC{sensN}{\fledbat\ vs \fledbat\: Sensitivity analysis of the fairness index to the number of flows, for different values of $\zeta$}
    \end{center}
\vspace{-4mm}
\end{figure}

Finally, we  present results for a varying number of \fledbat\ flows insisting on the bottleneck, with in $N\in[2,10]$. Every $k$-th flows arrive at  $t_k=10  k$\,s, and we evaluate the performance only after the $N$-th last flow has arrived in the bottleneck.
Results are reported in  \figR{sensN}, where we select a few values of $\zeta\in\{0.01,0.1, 0.2,0.5\}$ from the shaded gray zone of \figR{sensfLEDBAT}.  As it can be seen, it is always possible to find a value of  $\zeta$  that guarantees fairness for the whole set of flows, with any values in the range providing good results for the number of flows that are typically concurrently active in BitTorrent.  Moreover,  the very same values of  $\zeta$ that provide fair resource share, were already shown to provide efficient use of the resources for $N=2$ flows in  \figR{sensfLEDBAT}, which still holds when $N>2$ (omitted for lack of space).

\subsection{Realistic assessment of \fledbat\ vs \btledbat\ performance}

In order to compare \fledbat\ vs \btledbat\ performance of under more realistic conditions, we
finally consider a chunk-based scenario that loosely mimic the behavior of BitTorrent peers.
More specifically, the sender peer whose access bottleneck ($C=10$\,Mbps, $B=100$ packets)
is under observation opens a number $N=10$ of connections toward different destinations,
representing its neighbors in the peer swarm (whose downlink bandwidth in $\infty$ so that
bottleneck is at the access). Among the available connections, only a restricted number $M=5<N$
are concurrently active at any time, over which chunks of size  250\,kB ($approx$ 170 full size packets) are exchanged. The
sender peer uses only one protocol at a time, so that we compare two scenarios where flows
are either all \fledbat\ or \btledbat\ (i.e., intra-protocol case).
At the end of each chunk transmission, the sender chooses the next destination peer as follows:
with a \emph{persistence probability} $P_P$, the sender will send another chunk to the same
peer, keeping the congestion window settings; with probability $(1-P_P)$, the sender will
choose an inactive neighbor at random, resetting in this case the congestion window to 1.
Notice that, as $P_P\rightarrow 0$, we expect the performance of both protocol to be close:
indeed, when connections are reset every 170-th packet, the protocols are basically in transient
state, and the target is likely not even reached during the whole chunk transfer. Conversely,
differences are expected to arise in a more stable scenario $P_P\rightarrow 1$, where congestion
parameters are kept across chunks.

To add further realism, we consider both an \emph{homogeneous} network setup (i.e., in which
all peers have the same propagation delay $RTT=50$\,ms) as well as an \emph{heterogeneous} scenario
(i.e., in which each peer has a different propagation delay, which is the sum of 25\,ms on the 
forward path plus a delay on the backward path distributed according to realistic delay measurement 
effectuated by the Meridian project~\cite{meridian} with mean equal to 37.9\,ms).

%
%
\begin{figure}[t]
    \begin{center}
        \includegraphics[angle=-90,width=0.45\textwidth]{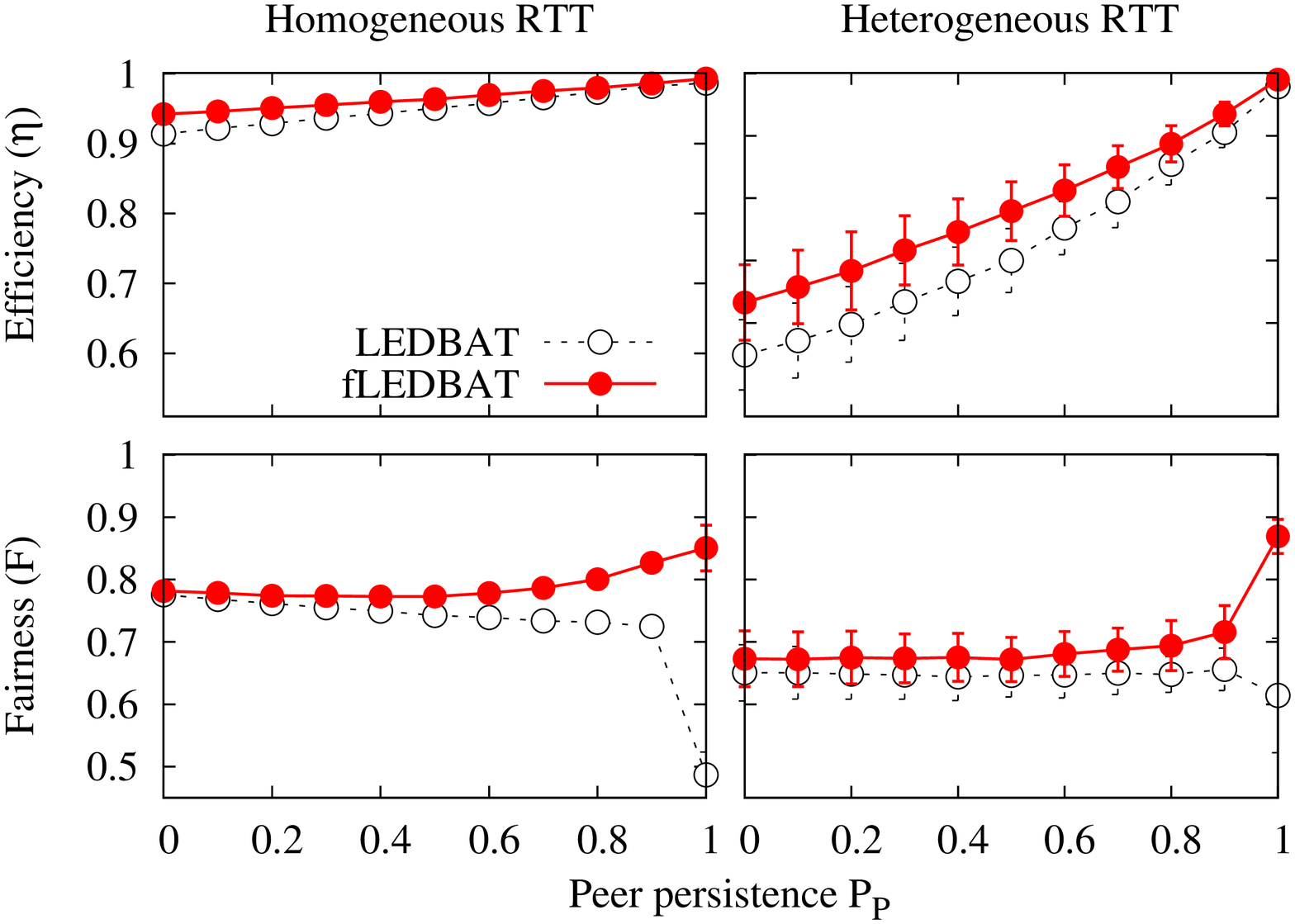}
        \figLC{realisticRTT}{Efficiency (top) and Fairness (bottom) of \fledbat\ in the homogeneous (left) and heterogeneous (right) RTT scenarios}
    \end{center}
\vspace{-4mm}
\end{figure}

Results of the comparison are reported, in term of efficiency and fairness, in
\figR{realisticRTT}, where we explore values of $P_P$ in the $[0,1]$ range and set $\zeta=0.1$
for \fledbat\ (which yielded good performance in the previous sections).
First of all, notice that, under all scenarios and $P_P$ values, \fledbat\ is more efficient
(due to AI) and fair (due to MD) with respect to \btledbat.
Likewise we expect BitTorrent to operate on $P_P\ge0.8$, where thus the gain from \fledbat\
adoption could be significant: indeed, BitTorrent try too keep the ``best'' (i.e., higher capacity)
peers  while at the same time trying to continuously discover new, potentially ``better'',
peers (i.e., by means of optimistic unchoking).

As far as fairness is concerned, consider the homogeneous case first, on the two plots on the left.
As expected, the performance gap exacerbates as $P_P\rightarrow 1$: in this case, \fledbat\ ability
to correctly measure the base delay leads to an increase of the fairness metric. On the contrary,
\btledbat\ fairness decreases as $P_P$ grow, due to the late-comer issue: the effect is stronger
when $P_P=1$, as in this case the unfair situation persists though the whole duration of the experiment
and leads to a consistent drop of $F$. Similar fairness considerations hold for the heterogeneous
case (see plots on the right), although in this case, due to $RTT$ heterogeneity, $RTT$ unfairness
arises in both \fledbat\ and \btledbat, reducing the absolute values achieved by the fairness index.

As far as efficiency $\eta$ is concerned, as expected, when the congestion window parameters are
reset every chunk ($P_P=0$), the link capacity is not fully utilized even in the homogeneous case.
The heterogeneous case further adds inefficiency, as flows with higher $RTT$ are also slower to
increase their congestion window, which result in more waste of link capacity.
However, it is worth pointing out that the additive increase component of \fledbat\ makes it more efficient than \btledbat\ under any circumstance, while the multiplicative decrease component guarantee at the same time its lower-priority with respect to TCP.

\section{Conclusion}\label{sec:outro}

In this work, we propose modification to the LEDBAT congestion control algorithm, that not only achieves \emph{low-priority inter-protocol} (i.e., against TCP) and \emph{efficiency intra-protocol} (e.g., with other \fledbat\  flows), but also reintroduce \emph{intra-protocol fairness}, solving thus the late-comer issues of the original \btledbat\ proposal.

We model the \fledbat\ dynamics via a fluid-model approach, which allows on the one hand to prove the correctness of the design, and that yield on the other hand closed-form expressions for the average rate and queue length.
By means of simulation, we further assess that \fledbat\ can safely operate under a number of scenarios (such as chunk-by-chunk and backlogged transmission) as it is not sensitive to the parameter selection, but operate reasonably well under a wide range of parameters. Overall, we see that our proposed modifications lead to \emph{provable performance} and constitute an \emph{improvement with respect to \btledbat} in terms of both fairness and efficiency. Our simulations confirm \fledbat\ robustness even under realistic heterogeneous network conditions (on which BitTorrent can be expected to operate).

\section*{Acknowledgments}
This work has been funded by the Celtic project TRANS.

\bibliographystyle{plain}
\bibliography{biblio}

\end{document}